\documentclass[usenatbib,]{mnras}

\usepackage[T1]{fontenc}
\usepackage{ae,aecompl}
\usepackage{graphicx}
\usepackage{epsfig}
\usepackage{natbib}
\usepackage{longtable}
\usepackage{times}
\usepackage{amsmath}

\newcommand{\nustar}{$Nu$STAR}
\newcommand{\sw}{$Swift$}
\newcommand{\xmm}{XMM-$Newton$}

\def \msun {$M_{\odot}$}
\def \apj {ApJ}
\def \apjl {ApJL}
\def \apjs {ApJS}\def \aap {A\&A}

\def \mnras {MNRAS}

\begin{document}

\title[Orbital period of IGR~J18214-1318]{Swift unveils the orbital period of IGR~J18214-1318}

\author[G. Cusumano et al.]{G.\ Cusumano$^{1}$, A.\ D'A\`i$^{1}$, A.\ Segreto $^{1}$, 
V.\ La Parola$^{1}$, M. Del Santo$^{1}$\\
$^{1}$INAF, Istituto di Astrofisica Spaziale e Fisica Cosmica,
Via U.\ La Malfa 153, I-90146 Palermo, Italy}

\date{}

\pagerange{\pageref{firstpage}--\pageref{lastpage}} \pubyear{}

\maketitle

\label{firstpage}

\begin{abstract}
We analysed 13 years of the Neil Gehrels Swift Observatory survey data collected on the High Mass X-ray Binary
IGR~J18214-1318. Performing the timing analysis we detected a periodic signal of $5.42$ d.
From the companion star characteristics we derived an average orbital separation
of $\sim 41 \rm R_{\odot}\simeq 2 R_{\star}$. The spectral type of the companion
star (O9) and the tight orbital separation suggest that IGR~J18214-1318 is a wind
accreting source with eccentricity lower than 0.17. The intensity profile folded
at the orbital period shows a deep minimum compatible with an eclipse of the source by the companion star.
In addition, we report on the broad-band 0.6--100 keV spectrum using data 
from \textit{XMM-Newton}, \textit{NuSTAR}, and \textit{Swift},
applying self-consistent physical models. We find that the spectrum is well 
fitted either by a pure thermal Comptonization component, or, assuming that the source 
is a neutron star accreting above the critical regime, by a combined thermal and 
bulk-motion Comptonization model. In both cases, the presence of a local neutral 
absorption (possibly related to the thick wind of the companion star) is required.
\end{abstract}

\begin{keywords}
X-rays: binaries -- X-rays: individual: IGR~J18214-1318. 

\noindent
Facility: {\it Swift}

\end{keywords}


        \section{Introduction\label{intro}}

During the last two decades astronomers have
taken advantage of two prolific
telescopes in the hard X-ray domain: the
IBIS/ISGRI telescope \citep{ubertini03,lebrun03}
on board the {\textit{International Gamma-Ray Astrophysics Laboratory}}
({\textit{INTEGRAL}}) satellite \citep{winkler03} and the
Burst Alert Telescope \citep[BAT][]{barthelmy05} on board
the {\textit{Neil Gehrels Swift Observatory}} (\citealp{gehrels04}; hereafter {\it Swift}). IBIS/ISGRI has
performed a deep and continuous scanning of the Galactic plane along the years
revealing a large number of new X-ray sources,
among which many were High Mass X-ray Binaries (HMXBs).
These are usually distinguished into two sub-groups
based on the observed spectral emission and variability:
obscured HMXBs and Supergiant Fast X-ray Transients
\citep[SFXT][]{sguera05, intzand05, negueruela06a, martinez17, bozzo17}.
The former are immersed in the wind from the companion star
and, as a consequence, strong absorption have made
their detection harder for soft X-ray instruments; the latter group shows very bright,
but rapidly transient flares, and were revealed thanks to the continuous
scan of the Galactic plane performed by {\textit{INTEGRAL}}. The association of
these sources to the class of HMXBs has been inferred either
through the discovery of their optical counterparts 
\citep[e.g.][]{filliatre04, chaty04, reig05, masetti06, negueruela06d, zurita08} or by the observation of long periodicities.
These can be due either to the occultation of the neutron star by the
supergiant companion or to the periodic enhancement of their X-ray emission
at the periastron passage of the neutron star in an eccentric orbit.
BAT is playing an important role in the study of many of these new {\textit{INTEGRAL}}
sources. Thanks to its large field of view (1.4 steradians
half coded) and to frequent changes in the satellite pointing
direction, BAT monitors daily $\sim$\,90 per cent of the sky,
making it an efficient tool to detect
transient phenomena from known and unknown sources \citep{krimm13}.
Combining the entire time span of its survey data, several
long periodicities of HMXBs have been revealed \citep[e.g.][]{corbet09, corbet10a,
corbet10b, corbet10c, corbet10d, corbet10e, cusumano10, laparola10, dai11a,
dai11b, laparola13, cusumano13a, cusumano13b, segreto13a,segreto13b,laparola14, dai15, cusumano15,
cusumano16}.

In this work we present a temporal and spectral analysis of
IGR~J18214-1318, a source discovered by {\textit{INTEGRAL}} on the
Galactic plane.
This source was observed  with a flux of $\sim$\,1 mCrab in the energy band
17--60 keV \citep{bird06, krivonos12, bird16} and
localised through a \textit{Chandra} observation at coordinates (J2000)
R.A.\,=\,18h21m19.76s, Dec.\,=\,-13$^{\circ} 18' 38.9''$
\citep{tomsick08}.
IGR~J18214-1318 is associated to USNO-B1.0 0766-0475700, most likely a O9I star,
and classified as an obscured HMXB \citep{butler09}.
The \textit{Chandra} spectrum is well modelled by a simple power law with a photon index $\Gamma$\,=\,$0.7^{+0.6}_{-0.5}$,
absorbed by an equivalent absorption column $N_{\mathrm{H}}$\,=\,(1.2$\pm$0.3)\,$\times$\,$10^{23}$ cm$^{-2}$.
Using \textit{Swift} data, \cite{rodriguez09}
measured a photon index of $\Gamma=0.4\pm0.2$ and a column density of
$N_{\mathrm{H}}$\,=\,$3.5^{+0.8}_{-0.5} \times 10^{22}$ cm$^{-2}$,
significantly lower than the value measured
with \textit{Chandra} and consistent with the Galactic $N_{\mathrm{H}}$
along the line of sight to IGR J18214-1318.
A high-statistics broadband spectrum from data collected by \textit{\xmm} \citep{jansen01}
and \textit{NuSTAR} \citep{harrison13} could be well modelled in the hard X-ray
region with a power-law
modified by an exponential cut-off with
e-folding energy $<25$ keV and a cut-off at $\sim$ 10 keV.
In the softer band, an equivalent fit could be obtained
either by adding a black-body component
with a temperature of $\sim$\,1.74 keV or with a partial covering
absorber of $\sim$\,10$^{23}$ cm$^{-2}$ and $\sim$77\% of
covering fraction \citep{fornasini17}.
In both cases, an iron K$\alpha$ emission line at
6.4 keV was detected with an equivalent width $\sim$\,55 eV.
Timing analysis did not reveal any
periodicity in the frequency range 0.1--88 Hz
with a 90\% upper limit on the rms noise level of 2.2\%.

This paper is organised as follows: Sect.~\ref{sect:data} describes the data reduction and
the calibration procedures applied to the data; in Sect.~\ref{sect:timing} and Sect.~\ref{sect:spectral} we
describe our timing and spectral analysis; in Sect.~\ref{sect:discussion} we discuss our results.

        \section{Data Reduction}\label{sect:data}

We made use for this work of data from BAT, XRT
\citep[X-ray Telescope][]{burrows04}, \textit{XMM-Newton}, and \textit{NuSTAR}.

We retrieved BAT survey data between 2004 December and  2017 February from the
HEASARC public
archive\footnote{\url{http://heasarc.gsfc.nasa.gov/docs/archive.html}}
and  processed them using the {\sc batimager} code \citep{segreto10}, dedicated to
the processing of coded mask instrument data.
IGR~J18214-1318 is detected with a significance of 24.7  standard deviations
in the 20--85 keV all sky map. 
For the timing analysis, we extracted a light curve in the same energy range
with the maximum available time resolution of $\sim$\,300 s and corrected to the Solar
System Barycentre (SSB) by using the task {\sc earth2sun}
and the JPL DE-200 ephemeris \citep{standish82}.
For the
spectral analysis, we produced
the background subtracted spectrum in eight energy channels, averaged over
the entire exposure, and we used the official BAT spectral redistribution
matrix.

XRT observed IGR~J18214-1318  four times.
The source was always observed in Photon Counting (PC)
mode \citep{hill04} for a total exposure of $\sim$\,9.4 ks.
The details on the XRT observations are reported in Table 1.
We processed the data using standard
filtering and screening criteria (0-12 grade selection, {\sc xrtpipeline}, v.0.12.4).
IGR~J18214-1318 was detected in 3 observations.
The source events were extracted from a circular region
(20 pixel radius, with 1 pixel = 2.36 arcsec) centred on the
source coordinates \citep{tomsick08}.
The background for the spectral analysis was extracted from an annular
region  with inner and outer radii 30 and 70 pixels, respectively.
XRT ancillary response file were generated with
{\sc xrtmkarf}; we used the spectral redistribution matrix v014.
For the spectral analysis we used only events from Obs.ID 00035354001 because of
its much higher signal-to-noise ratio with respect to the other observations.
XRT energy channels were binned requiring a minimum of 20 counts per bin in
order to use  the $\chi^2$ statistics.

\textit{NuSTAR} and \textit{\xmm}~ observed IGR J18214-1318 simultaneously on
2014 September 18. Details of these two observations are
reported in \citet{fornasini17} and summarised in Table~\ref{tab:tab1}.
We re-extracted data for spectral analysis using  NuSTARDAS v1.5.1
and the  Science Analysis Software (SAS) v15.5.0 for \nustar~ and
\xmm, respectively. We applied standard selection criteria and
source and background regions.  Spectral analysis was performed using {\sc xspec} v.12.5.
and spectral errors are given at 90\% confidence level.

\begin{table*}
\caption{Observations log. The quoted orbital phase refers to the
profile reported in the middle panel of Figure~\ref{period}. \label{log}}
\small
\begin{center}
\begin{tabular}{r l l l l l l l } \hline
Obs \# & Observatory & Instrument    & Obs ID &$T_{start}$  & Exposure    & Rate            & Orb.  Phase  \\
       &             &               & &  MJD        &  (ks)       &  (c/s) &         \\ \hline \hline
1      & \textit{Swift}    & XRT     &00035354001 & 53777.646   & 6.3         &$0.30\pm0.01$ &0.44     \\
2      & \textit{Swift}    & XRT     &00035354003 & 56240.844   & 0.7         &$0.15\pm0.01$ &0.52     \\
3      & \textit{Swift}    & XRT     &00035354005 & 57702.037   & 0.5         &$...$ &0.89     \\
4      & \textit{Swift}    & XRT     &00035354006 & 58064.671   & 0.8         &$0022\pm0.005$ &0.74     \\
\hline
5      & {\textit{NuSTAR}} &FPMA     & 3000114002 & 56918.107   & 26.3        &$0.56\pm0.01 $ &0.37     \\
       &                   &FPMB     & &             & 26.3        &$0.55\pm0.01 $ &         \\
\hline
6      &\textit{XMM-Newton}&EPIC--pn & 0741470201 & 56918.053   & 18.6        &$1.19\pm0.01 $ &0.36     \\
       &                   &EPIC--MOS1     & &             & 25.9        &$0.37\pm0.004$ &         \\
       &                   &EPIC--MOS2     & &             & 25.9        &$0.37\pm0.004$ &         \\
\hline \hline
\end{tabular}
\end{center}
\label{tab:tab1}
\end{table*}

        \section{Timing analysis}\label{sect:timing}

\begin{figure}
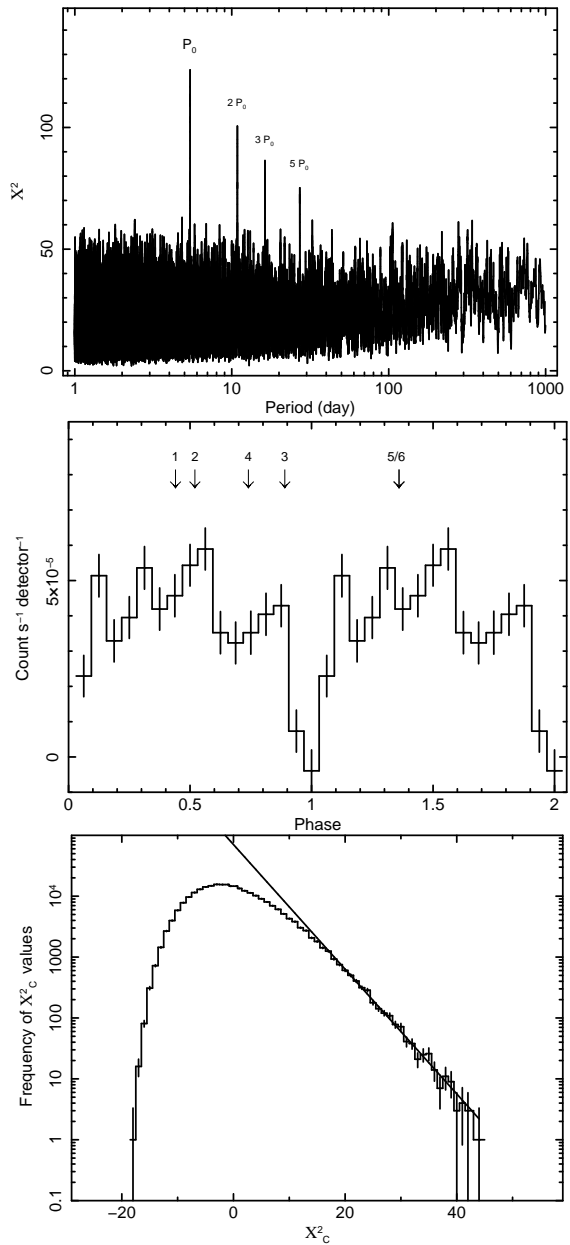

\begin{center}
\centerline{\includegraphics[width=5.4cm,angle=270]{CXOU_J182119_7-131838_fig3.eps}}
\centerline{\includegraphics[width=5.4cm,angle=270]{CXOU_J182119_7-131838_fig1.eps}}
\centerline{\includegraphics[width=5.4cm,angle=270]{CXOU_J182119_7-131838_fig2.eps}}
\caption[]{Top panel: periodogram of BAT (20--85\,keV)
data for IGR~J18214-1318.
Middle panel: Light curve folded at a period  P$_0= 5.4246$\,days, with 16 phase
bins. The arrows mark the orbital phase of the XRT observations (1 to 4)
and of the \textit{\xmm}, \textit{NuSTAR} observations (5/6).
Bottom panel: Histogram distribution of the $\chi^2_{c}$ values; the solid line is
the exponential function that best fits the right tail of the distribution.}
\label{period}
\end{center}
\vspace{-0.5truecm}
\end{figure}

We searched for periodicities in the 1--1000 d range
in the BAT survey data using the folding technique
and selecting events in the 20--85 keV energy range for optimal SNR.
The time resolution is given by $\Delta P=P^{2}/(N \,\Delta T)$, where P is the trial
period, $N=16$ is the number of phase bins used to build the trial profile,
and $\Delta T=$404.4 Ms is the data time span. The BAT survey data
present a large spread of statistical errors mainly due to the wide range of
off-axis directions in which the source is observed.
To overcome this issue, the rate in the folded profile for each trial period was
weighted by the inverse square of the corresponding statistical error
\citep{cusumano10}. The resulting periodogram (top panel in Fig.~\ref{period}) shows
several features emerging above the noise: the highest peak is
at P$_0=5.4246\pm0.0004$ d ($\chi^2\sim123$; the error is the period resolution
at P$_0$). The other peaks are multiples of P$_0$ (2, 3 and 5
times P$_0$). The intensity profile (middle panel in Fig.~\ref{period}) at P$_0$ with
T$_{\rm epoch}$\,=\,55684.71093750 MJD shows a flat intensity level and a deep minimum
with intensity consistent with no emission.
The  centroid of the minimum, evaluated by
fitting the data around the dip with a Gaussian model, is at a phase
0.987$\pm$0.010 corresponding to T$_{\rm min}$ = (55684.64 $\pm$ 0.05) $\pm$ nP$_0$
MJD. \\
The time variability of the source causes the average  $\chi^2$ in the
periodogram to significantly deviates from the average value expected for white
noise (N--1). As a consequence the $\chi^2$ statistics cannot be applied to
evaluate the significance of the detected periodicity.
Therefore, we determined the significance of the feature from the data in the
periodogram adopting the following methodology:
\begin{enumerate}
\item We fit the periodogram with a second-order polynomial;
a new $\chi^2_{c}$ periodogram was obtained by subtracting the best fit trend
from the original $\chi^2$ distribution. In the new periodogram the P$_0$ ha a
$\chi^2_{c}$ value of 102.8.
\item We build the histogram of the $\chi^2_{c}$ distribution
(Figure~\ref{period} bottom) selecting the values in the period interval between 1
and 10 d, excluding the values within an interval centred on P$_0$ and
$10\times\Delta P_0$ wide.
\item The tail ($\chi^2_{c} > 20)$ of the histogram is fitted with an
exponential function and its integral between 102.8 and infinity, normalised
for the total area below the histogram, is evaluated.
\end{enumerate}

The value we obtain (3.3\,$\times$\,10$^{-11}$) represents the probability of random
occurrence for $\chi^2_{c} > 102.8$  and corresponds to a significance of 6.6
standard deviations in Gaussian statistics.

The rate observed in the XRT observations
(Table~\ref{log} and middle panel in Figure~\ref{period}) shows a strong variability that 
cannot be explained with the shape of the BAT folded profile.
Observation 3, where the source is not
detected, is close to the dip of the pulse profile, while observation 4, that
shows a rate $\sim$\,10 times lower than observations 1 and 2, is far from the dip.

\section{Spectral analysis} \label{sect:spectral}
We re-analysed the data from simultaneous \textit{NuSTAR} and \textit{XMM-Newton}
observations performed in 2014, previously reported in \citet{fornasini17}. 
We aim at giving additional information on the the source by using physical models to explain the broadband X-ray emission. 
As a first check, we re-extracted the data and re-binned each spectrum according 
to the prescriptions outlined in \citet{kaastra16}\footnote{We used the ad-hoc script
written by C.\,Ferrigno at \url{https://gitlab.astro.unige.ch/ferrigno/optimal-binning}}. 
We applied the same models used by \citet{fornasini17} and obtained, within the statistical uncertainties, consistent parameters values. \citet{fornasini17} showed that the spectrum is well fitted by a  phenomenological model composed of a power-law with a high-energy cut-off; in addition, in the softer band, the spectrum needs either  a soft black-body  or a partial covering component, 
which were found statistically equivalent. It is known that the exponentially high-energy cut-off 
is an empirical model which suffers of artefacts  due to the discontinuity created
by the model at the cut-off energy. As discussed in \citet{fornasini17}, 
the spectral shape of IGR~J18214-1318 is compatible with the emission observed in accreting X-ray pulsars, 
even though a search for coherent pulsations 
did not reveal any periodic signal. In this scenario, the high-energy X-ray emission is
dominated by the emission from the shock in the accreting
column. The free-falling plasma is slowed down within few free path lengths 
by the presence of Coulomb, or radiative, shock
depending on the pulsar being in the critical regime, or not \citep{becker07}.  
In both cases, most of the hard X-ray radiation escapes either by bulk-motion 
or thermal Comptonization  processes in the post-shock region.
At high accretion rates, thermal Comptonization should be the 
dominant channel and pulsar spectra clearly show a cut-off at the 
electron thermal temperature superimposed on the hard power-law emission ($\Gamma$<2). 
At lower accretion rates, spectra appear softer and with higher, or absent, roll-over.
We first adopted a model of pure thermal Comptonization 
and then applied a self-consistent X-ray pulsar model, where all the main physical 
mechanisms are taken into account 
\citep[model \texttt{bwcycl}][]{becker07, ferrigno09} and then compared the results.

To model the thermal Comptonization we adopted the {\tt
nthcomp} model in \textsc{xspec} \citep{zdziarski96, Zycki99}. 
The soft seed photons with a black-body spectrum of temperature $kT_{\rm bb}$, 
produced in the NS polar cap or in the post-shock
region, are upscattered  by an electron population at temperature $kT_{\rm e}$ which is
related to the spectral high energy cut-off.
The model includes a fixed zero-width 6.4 keV line to fit the Fe
K{$\alpha$} emission and multiplicative factors for
each data-set to account for slight differences in the instrument intercalibration
(we fixed to 1 the FPMA constant, and set the EPIC/MOS1 and EPIC/MOS2 to be the same).
Line-of-sight interstellar absorption is modelled using the \texttt{tbabs} 
component, using cross-sections from \citet{verner96} and element abundances from \citet{wilms00}.
As in \citet{fornasini17}, we also found that residuals were present 
below 2 keV and the final fit result was 
not satisfactory ($\chi^2$/d.o.f.\,=\,465/428). Analogously, we added to this 
continuum model a black-body component, or, alternatively, a 
partial covering component. In the first scenario, we found a black-body 
temperature of 1.5\,$\pm$\,0.1 keV and a corresponding black-body radius 
of 0.4\,$\pm$\,0.1 km; the interstellar absorption column, $N_{\textrm{H, gal}}$,  was left free 
to vary and the best-fit value was (3.90\,$\pm$\,0.15) $\times$ 10$^{22}$ cm$^{-2}$ 
These values are compatible with the corresponding estimates reported 
in \citet{fornasini17}.
In the second scenario, we fixed the interstellar 
absorption to the Galactic expected value\footnote{We set this value 
according to the online $N_H$ estimator at 
\url{https://heasarc.gsfc.nasa.gov/cgi-bin/Tools/w3nh/w3nh.pl}} of 1.3\,$\times$\,10$^{22}$ cm$^{-2}$ \citep{HI4PI}
and found an excess of local absorption of (4.3$\pm$0.4)\,$\times$\,10$^{22}$ cm$^{-2}$ and an absorbed 
fraction of 89\,$\pm$\,2\%. However, unlike in \citet{fornasini17}, the partial 
covering model gave us a significantly better $\chi^2$ value  ($\chi^2$/d.o.f.\,=\,421/415) 
than that obtained by adding the black-body component  ($\chi^2$/d.o.f.\,=\,455/415). 
This statistical difference is mainly ascribed to the different spectral binning, 
because we noted a similar statistical difference for these two scenarios also
adopting the phenomenological continuum adopted in \citet{fornasini17}.

\begin{figure*}
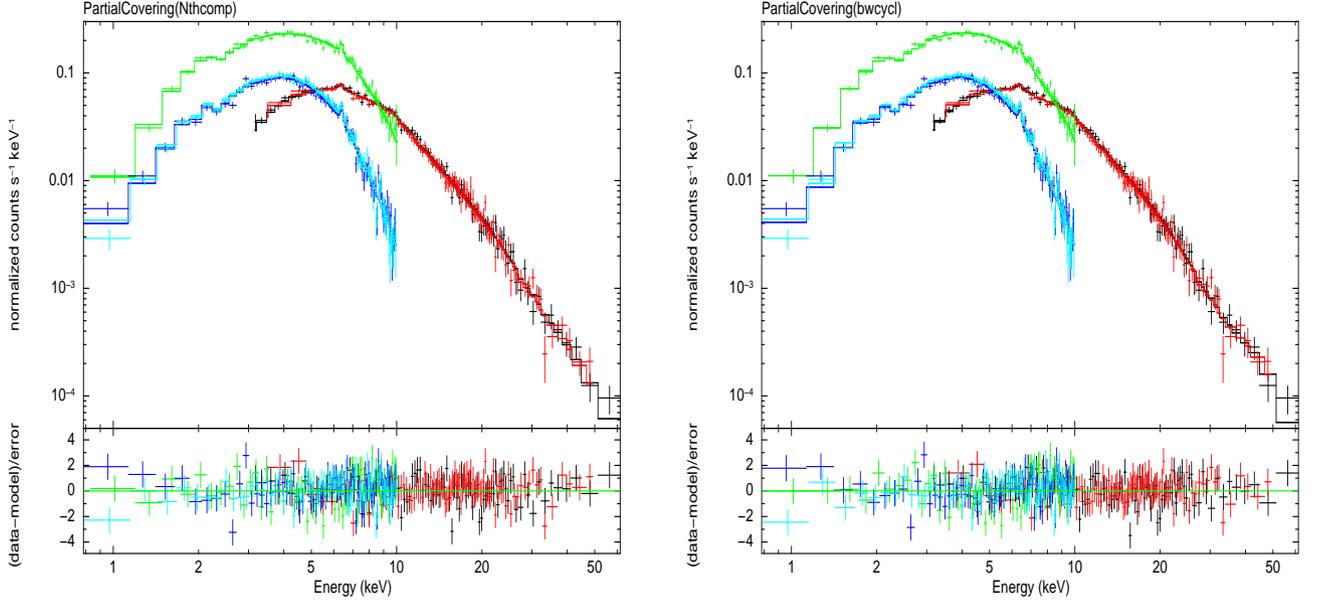

\begin{tabular}{c c}
\includegraphics[width=8cm, height=8.5cm, angle=-90]{plot_nthcomp.eps}     &  
\includegraphics[width=8cm, height=8.5cm, angle=-90]{plot_bwcycl.eps} \\
\end{tabular}
\caption{IGR J18214-1318 data and best-fitting models.
NuSTAR/FPMA and NuSTAR/FPMB data in black and red colours;
EPIC-PN data in red, EPIC/MOS1 and EPIC/MOS2 data in blue 
and cyan, respectively. Left panel:
data, best-fit model and residuals using the {\tt tbabs*pcfabs*(nthcomp+gau)}.
Right panel: data, best-fit model and residuals using the {\tt tabs*pcfabs*(bwcycl+gau)} model. 
Residuals in unit of standard deviations for the different data sets.} 
\label{fig:spectra}
\end{figure*}

It is worth noticing here that the fit sets only a poor constraint to the electron temperature,
with a lower limit of 13 keV (95\% confidence interval). Thus, we chose to fix it to a reference value of 20 keV,
since this is a typical value found in other accreting X-ray pulsars at similar luminosity \citep{coburn02}.
As a second step, we used the \texttt{bwcycl} model, assuming that the compact 
object is an accreting neutron star. This model has many parameters, most 
of which are strongly correlated, and it is important to fix as many of them as possible. 
In our context, we set to the default values the mass and the radius of the 
neutron star ($R_{\rm NS}$\,=\,10 km, $M_{\rm NS}$\,=\,1.4\msun), we assumed a distance of 10 kpc, a 
NS magnetic field of 4$\times$10$^{12}$ G.
From the \texttt{nthcomp} model, we derived a bolometric luminosity 
of  $\sim$\,10$^{36}$ erg s$^{-1}$, so we set for this model 
a mass accretion rate of 10$^{16}$ g s$^{-1}$. We left free to vary the following parameters: 
$\xi$, related to the escaping time of photons, $\delta$, related to the 
ratio of the bulk versus the thermal contribution of the whole Comptonized component, 
$r_0$, the radius of the accretion column and $T_e$, the temperature of 
the hot electrons \citep[see][for an extended discussion on the physical meaning of these parameters]{becker07}. 
This model gave a poor statistical fit to the data ($\chi^2$/d.o.f.\,=\,659/428), 
leaving a pattern of residuals reminiscent of the one obtained applying only the thermal 
Comptonization model. Again, we looked for the best-fit model adding either 
a black-body or a partial covering model, and, similarly to what obtained with 
the \texttt{nthcomp} model, we found a better description using the partial 
covering scenario (the $\chi^2$/d.o.f. is 506/415 and 425/415, for the black-body and 
partial covering scenarios, respectively). 
Since for both physical models we got better statistical 
results using a partial covering, and following \citet{fornasini17} who discussed 
the weakness of the black-body interpretation, hereafter we shall focus only on the partial covering scenario. 
We show in Fig. \ref{fig:spectra} the data,  
the best-fit models and residuals for the two models, and report in 
Table \ref{tab:tab2} the best-fitting parameter values and errors.
For both models, the iron line is well described by the same 
set of values: the energy is 6.39\,$\pm$\,0.03 keV, the line width 
is determined only as an upper limit of 85 eV (at 95\% confidence level); 
after freezing the width to zero, we derived a line normalisation 
of (1.7\,$\pm$\,0.4)\,$\times 10^{-5}$ photons cm$^{-2}$ s$^{-1}$ and 
a corresponding equivalent width of 54\,$\pm$\,2 eV.

\begin{table}
\begin{tabular}{ ll| rr}
\hline
Parameter & Units &  \multicolumn{2}{c}{Values}    \\ 
\hline \hline
          & &\texttt{nthcomp} & \texttt{bwcycl} \\ 
$N_{\textrm{H, gal}}$     & 10$^{22}$  cm$^{-2}$     & 1.3 & 1.3 \\
$N_{\textrm{H, pc}}$     & 10$^{22}$ cm$^{-2}$    & 4.3$\pm$0.4   & 9.6$_{-0.8}^{1.7}$  \\
$\rm F_{n_H,part}^{\textrm{a}}$ &                  & 0.89$\pm$0.02 & 0.76$_{-0.08}^{0.04}$ \\ \hline
$\Gamma$       &        & 2.07$\pm$0.03 & \\
$kT_{\textrm{bb}}$      & keV    & 1.39$\pm$0.04 & \\
$kT_{\textrm{e}}$       & keV    & 20 (fixed) & \\
$\xi$    & & & 2.07$\pm$0.17 \\
$\delta$ & & & 6.2$_{-1.4}^{3.0}$ \\
$T_e$    & keV & & 4.8$_{-0.2}^{0.1}$ \\
$R_o$    & m & & 5.5$\pm$0.3 \\
\hline
Flux$^{\textrm{b}}$ & & 6.0 &  6.1 \\ \hline
$\rm C_{FPMB}^{\textrm{c}}$   & & $1.04\pm0.02$ & $1.04\pm0.02$ \\
$\rm C_{PN}^{\textrm{c}}$     & & $0.80\pm0.01$ & $0.80\pm0.01$ \\
$\rm C_{MOS1/MOS2}^{\textrm{c}}$ & & $0.84\pm0.02$ & $0.84\pm0.02$  \\
\hline
$\chi^{2}$ / d.o.f. &     & 421/418 & 425/418 \\ 
\hline \hline
\end{tabular}
\caption{Best-fitting spectral parameters for the two 
models discussed in the paper.
$^{a}$Covering fraction for ${N_H, part}$. 
$^{b}$We report the unabsorbed 
(both for the Galactic and the local components) bolometric fluxes 
in the 0.1--100 keV with respect to the \textit{NuSTAR}/FPMA dataset 
in units of 10$^{-11}$ erg cm$^{-2}$ s$^{-1}$.
$^{c}$Multiplicative factors of the model for
each data-set. We used the reference constant of 1 for the 
\nustar\ FPMA dataset. 
\label{tab:tab2}}
\end{table}

\begin{figure}
\includegraphics[height=8cm, angle=-90]{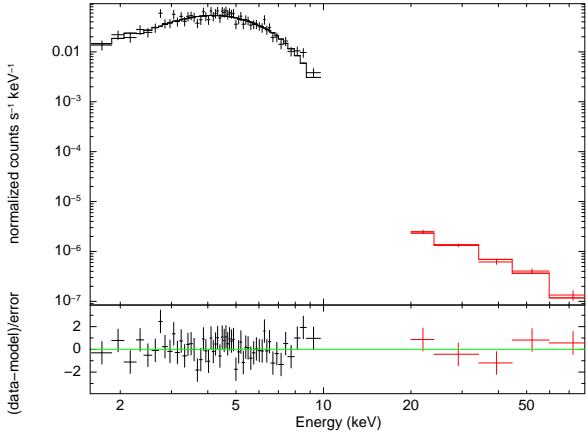}
\caption{XRT and BAT broad band spectrum of IGR J18214-1318. {\bf Top panel}:
data and best-fit model {\tt tbabs*pcfabs*(nthComp)}.
Bottom panel: residuals in units of standard deviations.}
\label{fig:xrtspectrum}
\end{figure}

Finally we also analysed the \textit{Swift} 
data, using the XRT data from ObsID 00035354001, for which there is the highest 
statistics, and the time-averaged BAT spectrum. The 1--10 keV X-ray 
spectrum is variable because of the changing of the local conditions on the 
neutral absorption and of the accretion rate  while the hard X-ray spectrum, above 15 keV, is generally 
dominated by the exponential tail of the Comptonized component, and depends only 
on the  electrons temperature and the instantaneous 
mass accretion rate. Assuming there is no significant change in the electrons 
temperature, we left a multiplicative constant free to vary  to keep into account 
the intercalibration between the two instruments and the different flux level. We find that  an 
absorbed thermal Comptonization gives a good description 
of the data ($\chi^2$/d.o.f.\,=\,48/60), though spectral parameters are not so well
constrained as in the previous case. We fixed the expected interstellar 
equivalent hydrogen column to 1.3\,$\times$\,10$^{22}$ cm$^{-2}$.
Using a partial covering we noted that the covering fraction parameter leaned to the higher 
boundary extreme, so that we could only measure a 
total absorption value of (4\,$\pm$\,1 )$\times$\,10$^{22}$ cm$^{-2}$; the electron
temperature resulted poorly constrained and fixed, then, to 20 keV; the soft seed-photon 
temperature, the $\Gamma$ parameter and the unabsorbed 0.1--10 keV flux are 
1.91\,$\pm$\,0.35 keV, 1.94\,$\pm$\,0.14
and (5.8\,$\pm$\,1.7)$\times$\,10$^{-11}$ erg cm$^{-2}$ s$^{-1}$, respectively. 
The multiplicative factor of the BAT model is 0.18\,$\pm$\,0.07, which indicates 
that the XRT observation caught the source in brighter state with respect
to long-term  averaged flux. The unabsorbed flux measured during the XRT observation 
results  a factor 2--3 higher than that observed in the simultaneous 
\textit{NuSTAR} and \textit{XMM-Newton} observations. 
We also found a similar amount of local absorption. 
In Fig.~\ref{fig:xrtspectrum}, we show data, best-fitting model and residuals 
for the combined XRT and BAT broadband spectrum.

Although we also obtained a satisfactorily description of the data
with the \texttt{bwcycl} model, we do not go into detail, as  slack 
constraints for many parameters prevented us to make 
meaningful comparisons or draw solid conclusions.\\

\begin{figure}
\begin{center}
\centerline{\includegraphics[width=9cm]{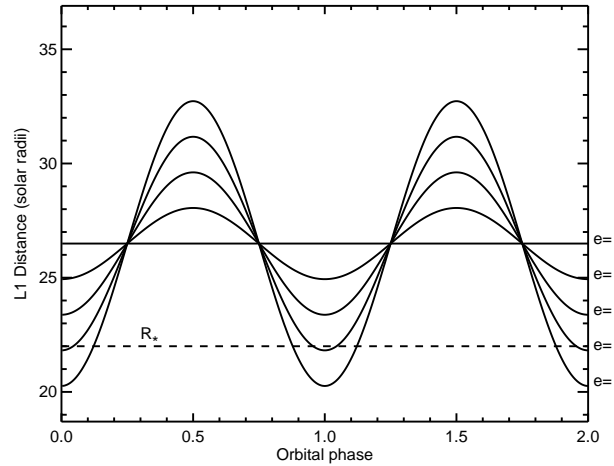}}
\caption[]{Distance of the L1 Lagrangian point from the companion star of
IGR~J18214-1318, as a function of the orbital phase, for different values of the
orbital eccentricity. The dashed line represents the radius of the companion
star.
}
                \label{lag}
        \end{center}
        \end{figure}

\section{Discussion}\label{sect:discussion}
We exploited archival data based on \sw, \xmm, and \nustar~ data available on
IGR J18214-1318 for an updated study of the spectral and timing properties of this source.
The BAT survey monitoring, spanning 13 years, reveals a periodic modulation
with P$_0$\,=\,5.4246\,$\pm$\,0.0004 d. The folded light curve shows a minimum
consistent with none, or negligible, emission, thus suggesting
the presence of an eclipse. We use Kepler's third law to derive the
semi-major axis of the binary system assuming that P$_0$ is the
system orbital period, $M_{\rm X}=1.4 M_{\odot}$
the mass of the neutron star, and $M_{\star} \simeq 30 M_{\odot} $ the
companion's star mass \citep{martins05}:
\begin{equation}
a = (G P_0^2~(M_{\star}+M_{\rm X})/4\pi^2)^{1/3} \simeq 41 R_{\odot}.
\end{equation}
Considering that the radius of the companion star is  $R_{\star}=\sim 22 R_{\odot} $
\citep{martins05}, the semi-major
axis length corresponds to $ \sim 2 R_{\star}$.
Such a tight orbital separation is common among wind-fed neutron stars accreting
from an O type companion star.  With this geometry, assuming the orbit to be nearly
edge-on, we expect the eclipse to last $\sim 16$\% of the orbit. This is roughly
consistent with the width of the eclipse observed in the folded light curve
(Fig.~\ref{period} middle).
The lack of detection in XRT observation 3, whose orbital phase falls marginally
outside the dip, could be explained with enhanced absorption of the soft
X-ray emission by the stellar wind, which results denser along the line-of-sight
for smaller angular separation.
However, we have also observed a significant flux variability in the soft X-rays, not related to the
orbital phase, so we cannot exclude that this non-detection is due to a flux
fluctuation because of a decrease of the accretion rate from the companion star.
Knowing the radius of the supergiant companion, we can estimate the upper limit
on the orbital eccentricity for a wind-fed accreting system. Fig.~\ref{lag}
shows how the Lagrangian point L1 varies with the orbital phase, for different
eccentricities \citep{paczynski71}. If the eccentricity were higher than
$\sim0.17$, L1 would be within the companion star radius, and the accretion would be
from Roche lobe overflow.

We have re-analyzed the broadband spectrum of IGR~J18214-1318 extending the spectral
analysis reported by \citet{fornasini17} by using physical models to fit the data.
A physical description is obtained either
by a pure thermal Comptonization model or by a more complex model which takes into account also 
the bulk-motion Comptonized component. In both cases, an excess below 
2 keV in the residuals is indicative of an additional component. 
\citet{fornasini17} explained this excess either with a partial covering 
or with the addition of a hot thermal black-body component, on the basis
of an equivalent statistical result. Our fits are instead significantly better when using 
the partial covering rather than the black-body. We found that the amount 
of local absorption can be uncertain by a factor of two depending on the 
choice of the continuum: the \texttt{bwcycl} model requires higher absorption values, 
similar to the results obtained by \citet{fornasini17}, the \texttt{nthcomp}
model requires half of this value and a higher covering fraction, close to 90\%, 
which suggests that local absorber embeds totally the compact object 
and reprocesses and re-emits the hard X-ray illuminating primary flux. 
The \texttt{bwcycl} model has been used under certain assumptions: 
that the compact object is a magnetized NS with a bipolar field of 
4\,$\times$\,10$^{12}$ G, and the accretion rate 
is close to the critical luminosity \citep{becker07}. These assumptions 
should be proved with future observations. The luminosity depends quadratically on 
the distance and on the estimate for the local absorption, and our best guesses  
at the moment favour a luminosity of $\sim$10$^{36}$ erg s$^{-1}$, which is expected below  
the critical luminosity. By adopting the appropriate transformations 
from our assumptions and from the best-fitting parameters, 
we derive the following physical quantities: the local mass accretion rate on the 
polar cap of the NS is of the order of 10$^{10}$ g cm$^{-2}$ s$^{-1}$, this 
builds a mound of material that has an altitude of $\sim$\,1 meter, a density 
of 13.5 g cm$^{-2}$ and a thermal temperature of 8.2$\times 10^{8}$ K \citep[Sect. 6.3 in][]{becker07}. 
The very low inferred radius of the accretion column leads to a lower 
critical luminosity compared to the standard bright X-ray pulsars 
of the order of a few 10$^{36}$ erg s$^{-1}$, which makes the adoption 
of this model reasonable \citep[see a similar 
discussion for the applicability of this model in the case of another 
accreting X-ray pulsar in][]{dai17}. Another important difference with other sources
examined using this model is the derived $\delta$ value that sets the 
relative importance of the  bulk vs. thermal Comptonization \citep[$\delta$\,=\,4 y$_{bulk}$/y$_{therm}$, where 
the y-parameter describes the fractional energy increase in each of these processes, see][]{becker07} . 
For this source $\delta$\,=\,6, which indicates that photons are mostly up-scattered 
by the free-fall electrons above the sonic point. 
Finally, we found that these models do also provide 
an adequate modelling to a \textit{Swift} observation
combined with an averaged long-term BAT spectrum,
though the lower statistics did not allow 
a tight comparison of these different observations.

\section*{Acknowledgments}

This work was supported by contract ASI I/004/11/0.
This work made use of data supplied by the UK Swift Science Data Centre at the
University of Leicester.
The authors acknowledge financial contribution
from the agreement ASI-INAF n.2017-14-H.0 and from the INAF mainstream grant (PI: T. Belloni).
%

\section*{Data availability}
The data underlying this article are available in the HEASARC archive at \url{https://heasarc.gsfc.nasa.gov/docs/archive.html} 
and, can be accessed with the following Dataset ID numbers: 30001140002 for \textit{NuSTAR}, 0741470201 for  \textit{XMM-Newton}, 
00035354001, 00035354003, 00035354005 and 00035354006 for Swift/XRT. 

\bibliographystyle{mnras}
\bibliography{refs}


\end{document}